\documentclass[%
 reprint,
 amsmath,amssymb,
 aps,
floatfix,
]{revtex4-2}

\usepackage{graphicx}
\usepackage{dcolumn}
\usepackage{bm}
\usepackage{physics}
\usepackage{xcolor} 
\usepackage[normalem]{ulem}
\usepackage{hyperref}
\usepackage{natbib}
\usepackage{amsmath}
\usepackage{comment}

\newif\ifcmnt
\cmnttrue 
\ifcmnt
    \providecommand{\aucmnt}[1]{#1}
    
\else
    \providecommand{\aucmnt}[1]{}
    
\fi

\begin{document}

\preprint{APS/123-QED}

\title{Quadrature Squeezing And Temperature Estimation From The Fock Distribution}

\author{Italo Pereira Bezerra}
 \email{italo@fisica.ufc.br}
\affiliation{Universidade Estadual do Ceará, Iguatu, Ceará, Brazil}
\affiliation{Departamento de Engenharia de Teleinformática, Universidade Federal do Ceará, Fortaleza, Ceará, Brazil}

\author{Hilma M. Vasconcelos}
 \email{hilma@ufc.br}
\affiliation{Departamento de Engenharia de Teleinformática, Universidade Federal do Ceará, Fortaleza, Ceará, Brazil\\
}

\author{Scott Glancy}
\email{scott.glancy@nist.gov}
\affiliation{National Institute Of Standards And Technology, Boulder, Colorado, USA}

\date{\today}

\begin{abstract}
We present a method to estimate the amount of squeezing and temperature of a single-mode Gaussian harmonic oscillator state based on the weighted least squares estimator applied to measured Fock state populations. Squeezing and temperature, or equivalently the quadrature variances, are essential parameters states used in various quantum communication and sensing protocols. They are often measured with homodyne-style detection, which requires a phase reference such as a local oscillator. Our method allows estimation without a phase reference, by using for example a photon-number-resolving detector. To evaluate the performance of our estimator, we simulated experiments with different values of squeezing and temperature. From 10,000 Fock measurement events we produced estimates for states whose fidelities to the true state are greater than 99.99\% for small squeezing ($r<1.0$), and for high squeezing ($r=2.5$) we obtain fidelities greater than 99.9\%. We also report confidence intervals and their coverage probabilities.
\end{abstract}

\maketitle


\section{\label{sec:level1} Introduction}

Estimating the state of a quantum system is an important tool for quantum information processing. It allows one, for example, to quantify the accuracy of prepared states, diagnose errors in the states, and estimate properties such as entanglement measures. State reconstruction is done in two steps: data extraction from the experiment and statistical estimation. For continuous variable systems, such as harmonic oscillators, the first is usually accomplished by using balanced homodyne detection to measure quadratures~\cite{lvovsky2005, PhysRevLett.106.220502, Raffaelli_2018}. Sets of quadratures measured at different phases can be used to reconstruct the system's Wigner function by using a numerical inverse Radon transform\cite{DTSmithey}, or one may reconstruct its Fock-basis density matrix by maximizing the likelihood function\cite{rehacek2007} or by other techniques. Another strategy for state estimation is to apply a displacement operator followed by Fock-state parity measurement, as done in Refs. \cite{Leibfried1, Hofheinz2009}. The homodyne-detection and the displacement-and-Fock-parity strategies require a phase reference to serve as a local oscillator and to apply the displacement, respectively. However, in some systems phase-sensitive detection is not easily available; examples are integrated optical circuits that use photon-number resolving detectors\cite{sahin2013, hopker2019} and trapped ions whose motional Fock states are measured by coupling to the ions' qubit states \cite{PhysRevLett.76.1796,scburd}. In such systems, one would still like to estimate those features of a quantum state that are phase-independent.

We focus on single-mode, Gaussian states centered at the origin of quadrature space, which we call ``squeezed thermal states''. Squeezed states have been studied for applications such as measurement noise reduction\cite{acernese2019, tse2019}, as generators of entanglement in continuous variable quantum teleportation \cite{furusawa1998}, and as resource states for quantum computation\cite{gottesman2001}. We present here a method for inferring a system's squeezing and temperature based on Fock state measurements, minimizing the weighted squared errors between the probabilities of measuring each Fock number and the frequency of observing that Fock number.  Because we measure only Fock populations, we do not have access to phase information. The method can be applied to any type of quantum oscillator such as superconducting resonators, single trapped ions, and photons. From 10,000 measurement events, we obtained fidelities between the true state and the estimated state greater than $99.99\%$ for small squeezing ($r<1.0$). Even for high squeezing ($r=2.5$) the average of the estimate's fidelity with the true state was greater than $99.9\%$.


The paper is organized as follows: in Section \ref{sec:level2} we present the Fock probability distribution and the weighted least squares estimator of squeezing and temperature, including our use of Bayes rule to obtain the weights. In Section \ref{sec:testing1} we report fidelities of estimated states obtained from simulated experiments, for different values of total number of measurements, temperature and squeezing. We discuss confidence intervals and bias correction in Section \ref{sec:testing2}, and make our concluding remarks in Section \ref{sec:conclusion}.

\section{\label{sec:level2} Fitting Model}

\subsection{Fock State Probability Distribution}
Squeezed thermal states can be obtained by different processes. Thermal noise can be added to a squeezed state, squeezing can be applied to a thermal state, or both heating and squeezing can happen simultaneously. All the resulting states are Gaussian and can be described by the parameters $\bar{n}$ (the mean thermal Fock number, which quantifies the temperature) and $r$ (the strength of the squeezing)\cite{olivares_s}.

Let us consider the case where squeezing is applied to a thermal state. The thermal state has the density operator
\begin{equation}
\hat{{\rho}}=\sum_{n} \frac{\bar{n}^n}{(\bar{n}+1)^{n+1}}\ket{n}\bra{n}.
\end{equation}
The state is then squeezed, resulting in the new state 
\begin{equation}
\hat{\rho}_s=\hat{S}^{\dagger}\hat{{\rho}}\hat{S},
\end{equation}
where $\hat{S} = \exp[\frac{r{\hat{a}}^2}{2}-\frac{r(\hat{a}^{\dagger})^2}{2}]$ is the squeezing operator, with a squeezing parameter $r$, and $\hat{a}$ and $\hat{a}^{\dagger}$ are the harmonic oscillator ladder operators. The quadrature operators are 
\begin{equation}
\begin{split}
   \hat{q}=\frac{(\hat{a}+\hat{a}^\dagger)}{\sqrt{2}},\\
   \hat{p}=\frac{(\hat{a}-\hat{a}^\dagger)}{i\sqrt{2}}. 
\end{split}
\end{equation}
Squeezed thermal states and coherent states are examples of Gaussian states, meaning that their Wigner distributions are Gaussian functions.  The most general form for a single-mode Gaussian state is given by the two-dimensional Wigner function:
\begin{eqnarray}
   W(q,p) = \frac{1}{2\pi|\Sigma|^{-1}}\exp \left[-\frac{1}{2}(X-X_0)^T\Sigma^{-1}(X-X_0)\right], \nonumber \\
\end{eqnarray}
where $X=(\hat{q}, \hat{p})^\text{T}$ is the quadrature vector, $X_0=\langle X \rangle$ is the displacement vector, and $\Sigma$ is the $2 \times 2$ covariance matrix of the Gaussian state. The thermal and squeezing information is encoded in the covariance matrix.

Fock state population measurements have no phase dependence. Therefore we cannot obtain a full estimate of $\Sigma$ because all phase rotations of $\Sigma$ give the same Fock distribution.  To remove this ambiguity, we estimate a diagonal $\Sigma$ with the quadrature variances $V_q= \langle \hat{q}^2 \rangle - \langle \hat{q} \rangle^2 \leq V_p =\langle \hat{p}^2 \rangle - \langle \hat{p} \rangle^2$:
\begin{equation}
\Sigma = 
\begin{pmatrix}
V_q & 0\\
0 & V_p
\end{pmatrix}.
\end{equation}
For a squeezed thermal state, the variances are related to the squeezing and temperature by
\begin{equation}
\begin{split}
    &V_q = \frac{1}{2}(2\bar{n}+1)e^{-2r},\\
    &V_p = \frac{1}{2}(2\bar{n}+1)e^{2r}.
\end{split}
\end{equation}

The experiment provides a list with the number of occurrences of each Fock state obtained from $N$ measurements. The probability of finding the system in a specific Fock number $n$ is
\begin{equation}
    P(n|V_q,V_p) = \text{Tr}\{\hat{\rho} \ket{n}\bra{n}\},
\end{equation}
where $\hat{\rho}$ is the density operator of the state. This probability could be calculated by overlapping the unknown state's Wigner function $W(q,p)$ and the $n$\textsuperscript{th} Fock state's Wigner function $W_n(q,p)$:
\begin{equation}
\label{directWigner}
    P(n|V_q,V_p) = 2\pi\iint_{-\infty}^{\infty}  W(q,p) W_n(q,p) dq \,dp,
\end{equation}
 where $W_n(q,p)$ is 
\begin{equation}
    W_n(q,p) = \frac{(-1)^n}{\pi}e^{-q^2-p^2}L_n (2q^2,2p^2),
\end{equation}
and $L_n (2q^2,2p^2)$ are the Laguerre polynomials.

$P(n|V_q,V_p)$ can also be obtained using the similar approach described in Ref.~\cite{PhysRevA.49.2993}: by calculating the diagonal elements of the density matrix in the coherent state basis, using the overlap relation
\begin{equation}
   \bra{\alpha}\hat{\rho} \ket{\beta}= \frac{1}{2\pi}\iint_{-\infty}^{\infty}   W(q,p)W_{\beta\alpha}(q,p) dq\,dp,
     \label{EQPnDodonov1}
\end{equation}
where $\ket{\alpha}$ and $\ket{\beta}$ are coherent states and $W_{\beta\alpha}$ is the Wigner function of $\ket{\beta}\bra{\alpha}$.

The result of the integral for squeezed thermal states is expressed, in terms of 2D Hermite polynomials $ H_{mn}^{\{R\}}({0,0})$, as:
\begin{equation}
\begin{split}
\bra{\alpha}\hat{\rho} \ket{\beta} = P&(0|V_q,V_p) \exp \left[ \frac{-(|\alpha|^2 + |\beta|^2)}{2} \right] \\ \times &\sum_{m,n=0}^{\infty}\frac{{\alpha}^n{\beta^*}^m}{m!\,n!}H_{mn}^{\{\boldsymbol{R}\}}({0,0}),
\end{split}
\label{wigner_overlap}
\end{equation}
where $\boldsymbol{R}$ is a symmetric $2 \times 2$ matrix whose elements  are
\begin{equation}
\begin{split}
&R_{11}=R_{22}=\frac{2(V_p-V_q)}{1+2(V_q+V_p) + 4(V_qV_p)}\\
&R_{12}=R_{21}=\frac{1-4V_pV_q}{1+2(V_q+V_p) + 4(V_qV_p)}.
\end{split}
\end{equation}

One can also compute $\langle \alpha | \hat{rho} | \beta \rangle$ by expanding each of the coherent states in the Fock basis to obtain
\begin{equation}
 \bra{\alpha}\hat{\rho} \ket{\beta} = \exp \left[ \frac{-(|\alpha|^2 + |\beta|^2)}{2} \right] \sum_{m,n=0}^{\infty}\frac{{\alpha}^n{\beta^*}^m}{(m!\, n!)^{\frac{1}{2}}}\rho_{mn}.
 \label{fock_expansion}
\end{equation}
By comparing Eq. (\ref{wigner_overlap}) and Eq. (\ref{fock_expansion}), ref. \cite{PhysRevA.49.2993} derives the expression for the diagonal density matrix elements $\rho_{nn}=P(n|V_q,V_p)$:
\begin{equation}
P(n|V_q,V_p) = P(0|V_q,V_p) \frac{H_{nn}^{\{R\}}(0,0)}{n!},
\label{EQPnDodonov2}
\end{equation}
where
\begin{equation}
{P}(0|V_q,V_p) = [0.25 + V_p V_q + 0.5\,(V_p+V_q)]^{-1/2},
\label{EQPnzero}
\end{equation}
\begin{eqnarray}
H_{nn}^{\{R\}}(0,0) ={n!}\left(\frac{[0.5 + 2V_p V_q - (V_p+V_q)]}{[0.5 + 2V_p V_q + (V_p+V_q)]}\right)^{n/2}  Q_{n}(f), \nonumber \\
\label{EQHermite0}
\end{eqnarray}
$Q_{n}(f)$ is the Legendre polynomial of order $n$, and
\begin{equation}
f =\frac{-(1-4V_q V_p)}{[(4V_qV_p +1)^2-4(V_q+Vp)^2]^{1/2}}.
\label{EQlegendre1}
\end{equation}
The function $P(n|V_q,V_p)$ obtained by the method of Ref.~\cite{PhysRevA.49.2993} was faster to calculate than using Eq.~(\ref{directWigner}), though the two methods give equal probabilities. Therefore, we use the probability distribution obtained by Eq.~(\ref{EQPnDodonov2}) in our code.

\subsection{Fitting Method}
The experiment measures the number $k_n$ of times each Fock state $n$ over a total of $N$ measurements, and we immediately obtain estimates of $P(n|V_q,V_p)$ by calculating $f_n=k_n/N$.  To estimate $V_q$ and $V_p$ we apply the weighted least squares estimator.
The weighted sum of squared residuals is given by:
\begin{equation}
    \Delta(V_q,V_p) = \sum_{n=0}^{n_F} {w_n} \left[P(n|V_q,V_p)- f_n\right]^2,
\end{equation}
where $w_n$ is the weight associated with the $n$\textsuperscript{th} measurement. Our estimates will be those values of $V_q$ and $V_p$ that minimize $\Delta(V_q,V_p)$. However, not all pairs of variances are allowed because the Heisenberg restriction imposes $V_q \times V_p \geq 0.25$. We also impose the constraints $V_q \leq V_p$, $V_q > 0$, and $V_p > 0$. 

The weights $w_n=1/\text{Var}(f_n)$ quantify the uncertainty in the measurement $f_n$, which we calculate as described below. We model the event of obtaining a specific Fock state $n$, $k_n$ times when we perform $N$ experiments with the binomial distribution.  The binomial distribution has success probability $p_n$, $f_n$ is an estimate of $p_n$, and $\text{Var}(k_n) = Np_n(1-p_n)$. A direct choice to estimate $\text{Var}(f_n)$ would be to use the maximum likelihood strategy. The maximum likelihood estimator for the binomial success probability $p_n$ is $\hat{p}_n = f_n = k_n/N$. We can estimate $\text{Var}(f_n)$ with
\begin{align}
    \text{Var}(f_n) & = \frac{\text{Var}(k_n)}{N^2} \\
    & \approx \frac{N\hat{p}_n(1-\hat{p}_n)}{N^2} \\
    & = \frac{k_n(N-k_n)}{N^3}.
\end{align}
With this estimate of $\text{Var}(f_n)$, the weight depends on $1/k_n$, and that will be a problem when $k_n = 0$, which will happen frequently with highly pure squeezed states.

To overcome this problem, we instead use Bayesian inference~\cite{bayes1} to estimate $\text{Var}(f_n)$. The posterior distribution of $p_n$ is given by
\begin{equation}
P(p_n|k_n)=P(k_n|p_n)P(p_n)/P(k_n).
\label{eq_bayes1}
\end{equation}
The likelihood $P(k_n|p_n)$ can be easily obtained by using the binomial distribution:
\[
P(k_n|p_n)=\binom{N}{k_n} {p_n}^{k_n} \,(1-p_n)^{N-k_n}.
\]
We choose the prior distribution $P(p_n)$ to be a Beta distribution since it is a conjugate prior for binomial distributions, meaning that the posterior $P(p_n|k_n)$ is also a Beta distribution. The prior is given by
\begin{equation}
P(p_n)= \frac{{p_n}^{\nu-1} (1-p_n)^{\eta-1}}{\text{Beta}(\nu,\eta)},    
\label{eq_prior}
\end{equation}
where $\text{Beta}(\nu,\eta)$ is the Beta function:
\[
\text{Beta}(\nu,\eta) = \int_{0}^{1}t^{\nu-1} (1-t)^{\eta-1}dt,
\]
and  \(\nu\) and \(\eta\) are the ``shape parameters''. For $P(k_n)$ we get 
\begin{equation}
\begin{split}
P(k_n)&=\int_{0}^{1}P(k_n|p_n) {P(p_n)}\,d{p_n}
\\&=\binom{N}{k_n}\frac{\int_{0}^{1}{p_n}^{k_n+\nu-1} \,(1-p_n)^{N-k_n+\eta-1}\,d{p_n}}{\text{Beta}(\nu,\eta)}\\
&=\binom{N}{k_n} \frac{\text{Beta}(k_n+\nu,N+\eta-k_n)}{\text{Beta}(\nu,\eta)}.
\end{split}
\label{eq_bayesnorm}
\end{equation}
The posterior distribution is then given by
\begin{equation}
\begin{split}
P(p_n|k_n)=\frac{{p_n}^{k_n+\nu-1} \,(1-p_n)^{N-k_n+\eta-1}}{\text{Beta}(k_n+\nu,N+\eta-k_n)},
\end{split}
\label{eq_posterior}
\end{equation}
which is a Beta distribution with new shape parameters \(\nu'=k_n+\nu\) and \(\eta'=N+\eta-k_n\) and variance
\begin{equation}
    \text{Var}(p_n|k_n)=\frac{(k_n+\nu) (N+\eta-k_n)}{(\nu+N+\eta)^2 (\nu+N+\eta+1)}.
\end{equation}
We use this variance when computing the weights $w_n=1/\text{Var}(p_n|k_n)$ in the weighted sum of squared residuals, which ensures that the weights are finite when $k_n=0$.

\begin{figure}
    \centering
    \includegraphics[scale=0.36]{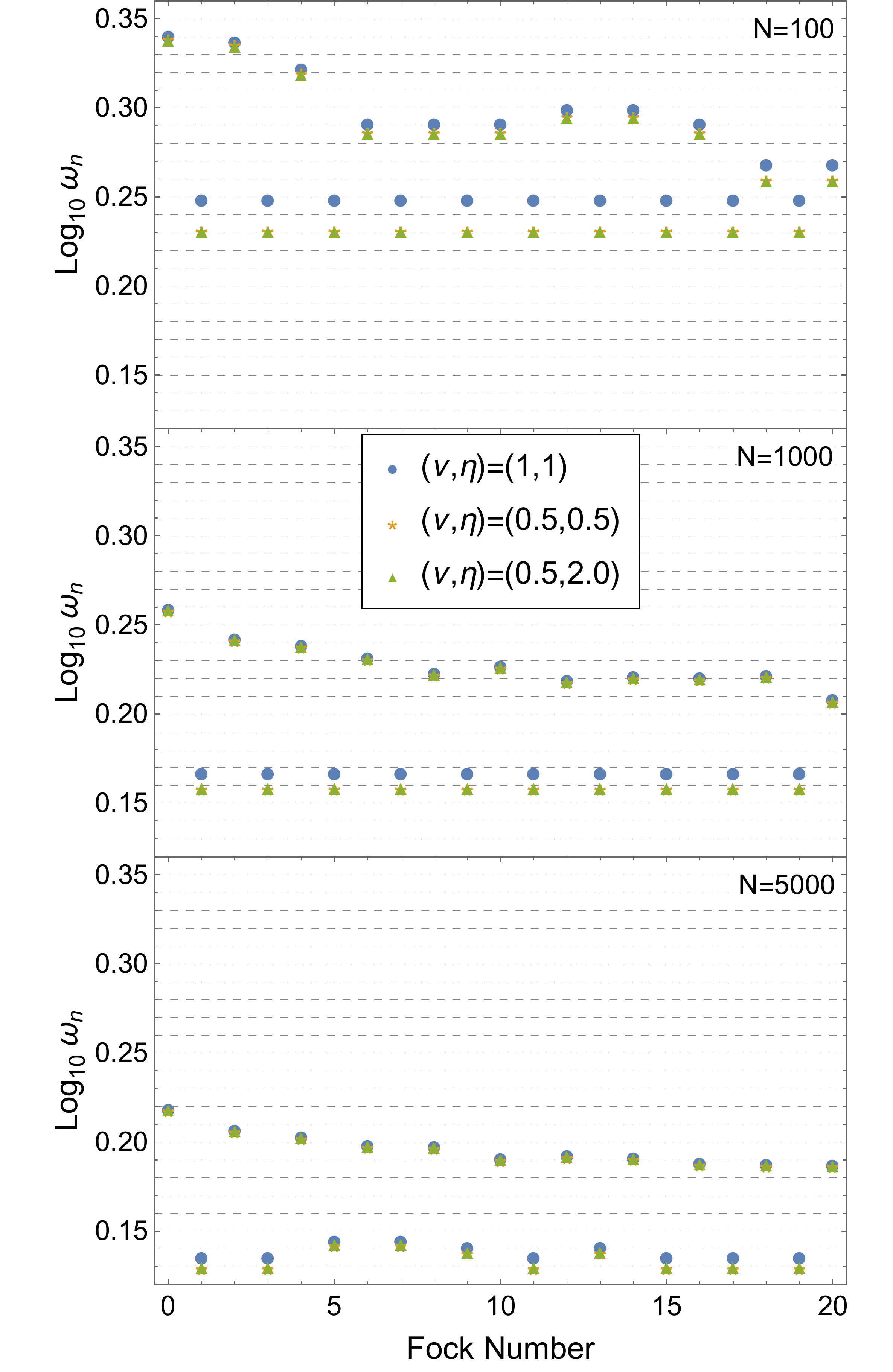}
    \caption{Graphs showing the logarithm $\log_{10}{w_n}$ of the weights used in the least-squares fit versus $n$ for measurements of the quantum state with squeezing $r=2.5$ and thermal average Fock number $\bar{n}=0.01$ for three different numbers $N$ of measurements and for different values of the prior distribution's shape parameters $\nu$ and $\eta$. We see that the weights are not very sensitive to the prior distribution's shape parameters $\mu$ and $\nu$, and the closeness increases for increasing $N$. We observed similar behavior for other choices of $r$ and $\bar{n}$. }
    \label{fig:Wshapeparams}
\end{figure}

Fig.~\ref{fig:Wshapeparams} presents the behavior of weights calculated with priors having different values of $N$ and different shape values. The graphs show that for $N\geq 100$, which is easily achievable in experiments, the weights are not very sensitive to the choices for $\eta$ and $\nu$. Odd Fock numbers have higher weights because we are testing a nearly pure squeezed state, which has low probability for containing odd Fock numbers. 

\section{Testing}
\label{sec:testing}

\subsection{Simulated Experiments}
\label{sec:testing1}

To test our estimator we feed it with simulated data from a known state and compare it with the output. We also present estimates of bias, confidence intervals, and their coverage probabilities. For various choices for $\Bar{n}$ and $r$ we simulated 100 experiments. For each experiment, the Fock distribution is measured $N$ times and an estimate of $\Bar{n}$ and $r$ is produced.
From those experiments, we calculated the mean fidelity and report its dependence on $N$ for $\Bar{n} =( 0.001, 0.01, 0.1, 2)$ and $r = (0, 1.0, 2.5)$. (Ref. \cite{scburd} reported squeezing of $r=2.26 \pm 0.02$.)

Of course only a finite number of Fock states can be resolved in an experiment. Ref. \cite{scburd} reported the resolution of 20 Fock states for ion motion, and ref. \cite{lolli2012} and ref. \cite{morais2020} resolve 29 and 16 photonic Fock states, respectively, with transition edge sensors. In our simulations, we assume that the detectors can resolve Fock numbers 0 through 20, but they cannot distinguish higher Fock numbers. Thus we have 22 possible measurement results, with the first 21 being Fock numbers 0 through 20 and the last containing all events with Fock numbers $\geq 21$. This provides enough information for our tests, which have $r \leq 2.5$ and $\bar{n} \leq 2$. During the simulated experiment the probability of getting a set of counts for $n = 0$ to 21 over a total of $N$ measurements can be calculated by a multinomial distribution for $P(n|V_q,V_p)$ and $N$ total measurements. 
We used Mathematica to generate samples from a multinomial distribution. Thus we simulate all the data needed to evaluate our method.

Quantum state fidelity $F$ measures the closeness of two states $\rho_1$ and $\rho_2$:
\begin{equation}
F(\rho_1,\rho_2) = \left[\text{Tr}\left(\sqrt{\sqrt{\rho_1}\rho_2\sqrt{\rho_1}}\right)\right]^2.
\end{equation}
We use the quantum state fidelity between the true state and our estimate to quantify the accuracy of our estimator. Because we are estimating squeezed thermal states, we rewrite the fidelity as~\cite{MarianMarian}:
\begin{equation}
\begin{split}
F(\rho_1,\rho_2) &=(\sqrt{\Xi+\Lambda}+\sqrt{\lambda})^{-1},
\end{split}
\end{equation}
where $\Xi$ and $\Lambda$ are given by
\begin{equation}
\begin{split}
&\Xi = \text{det}(\Sigma_1 + \Sigma_2),\\
&\Lambda = 4\, \text{det}\left(\Sigma_1 + \frac{i}{2}J\right)\, \text{det}\left(\Sigma_2 + \frac{i}{2}J\right),\\
\end{split}
\end{equation}
$\Sigma_1$ and $\Sigma_2$ are the single mode covariance matrices for each mode, and $J$ is 
\begin{equation}
J = 
\begin{pmatrix}
0 & 1\\
-1 & 0
\end{pmatrix}.
\end{equation}

\begin{figure}
    \centering
    \includegraphics[scale=0.35]{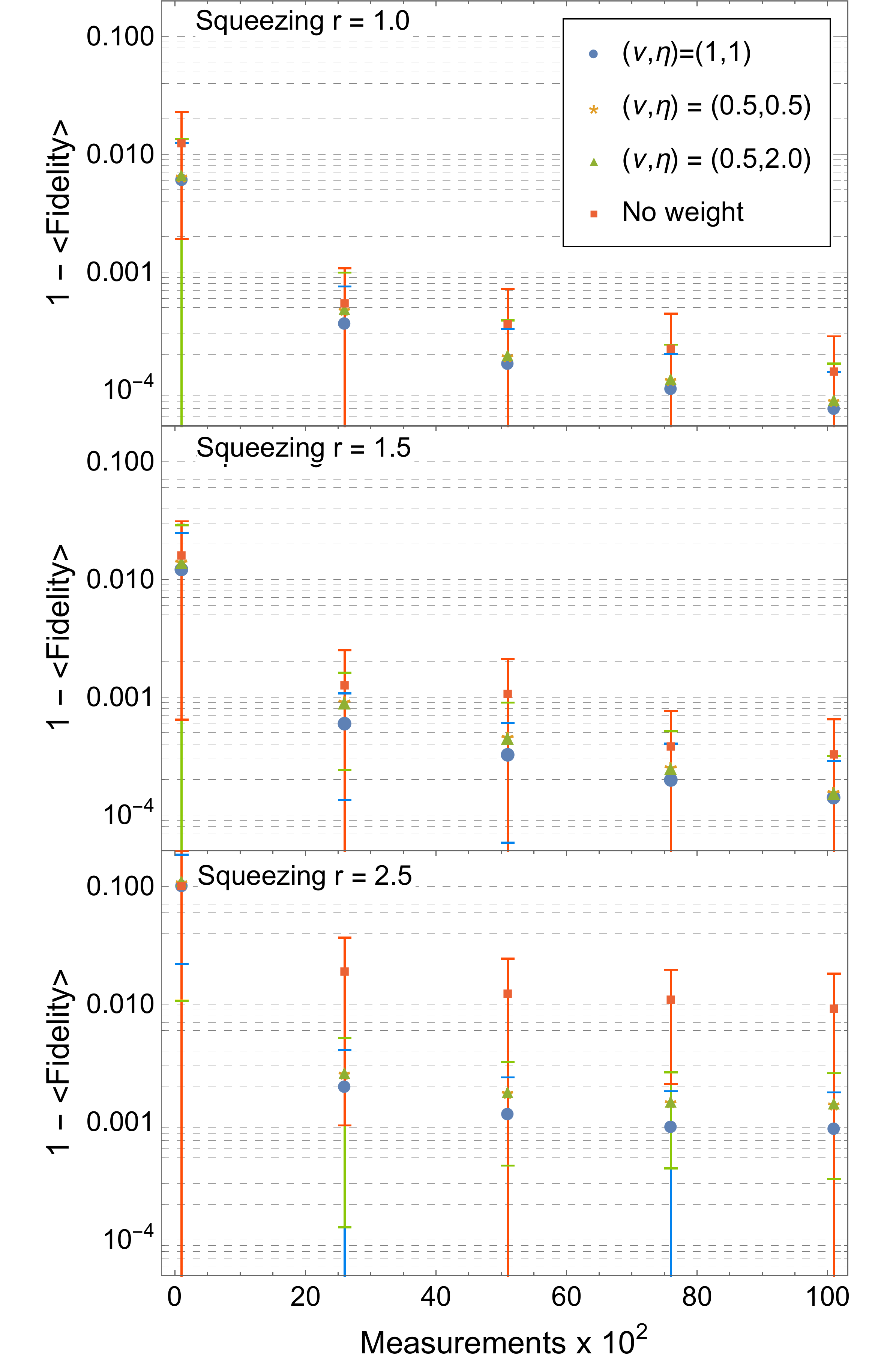}
    \caption{$1 - \langle\text{Fidelity}\rangle$, the mean infidelity averaged over 100 simulated experiments, and the standard deviation of the 100 infidelities as a function of the number $N$ of measurements for $\bar{n}=0.01$, three values of squeezing, and different choices the shape parameters $\nu$ and $\eta$ that determine the weights used in the weighted least squares estimator.}
    \label{fig:Wconparisons}
\end{figure}

Fig.~\ref{fig:Wconparisons} plots the mean infidelity, $1-\langle\text{Fidelity}\rangle$, averaged over 100 simulated experiments using different shape parameters $\mu$ and $\nu$ (used for choosing the weights) on the fidelities as a function of $N$ for three different states. For any $N$ and the three states chosen, $\mu=\nu=1$ (corresponding to the uniform prior distribution) performs better than the other tested pairs of shape parameters. For all following simulations we use $\mu=\nu=1$. Fig.~\ref{fig:Wconparisons} also shows the importance of using weights, specially for high squeezing states ($r=2.5$).

\begin{figure}
    \centering
    \includegraphics[scale=0.293]{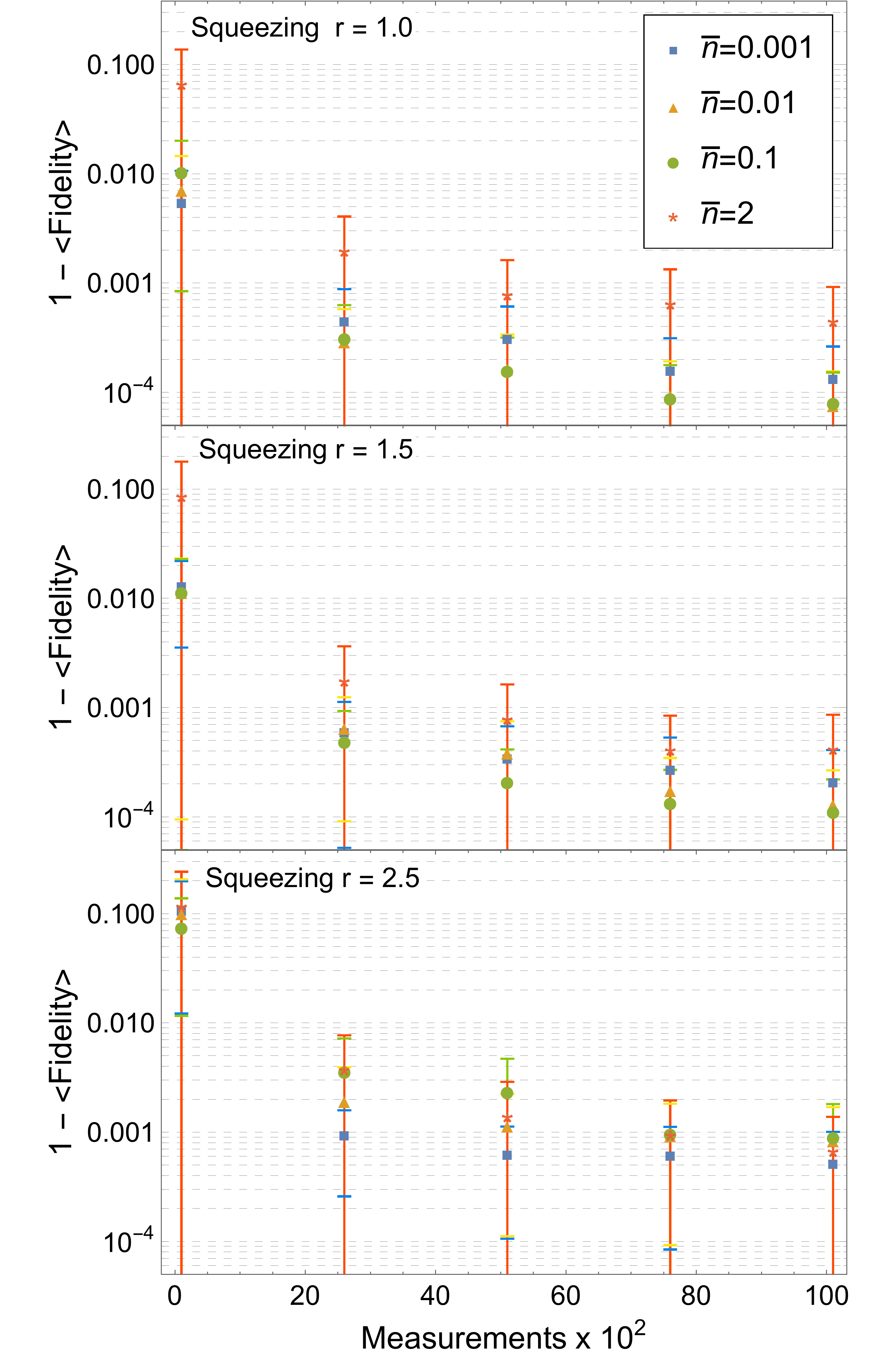}
    \caption{$1-\langle\text{Fidelity}\rangle$ and its the standard deviation of the fidelity estimates as a function of the number of measurements for different $\bar{n}$'s.}
    \label{fig:FxN_1}
\end{figure}

Fig. \ref{fig:FxN_1} presents $1 - \langle \text{Fidelity}\rangle$ versus $N$ for three different squeezing values and four values of $\bar{n}$. States with higher squeezing require larger $N$ to obtain high fidelity estimates, but the fidelities are less sensitive to $\bar{n}$. For a state with $r=2.5$, $\bar{n}=0.1$, after $10100$ measurements, from 100 simulated experiments, we obtain an average fidelity of $0.9991$, and the standard deviation of the fidelity estimates is $0.0011$.

\subsection{Confidence Intervals}
\label{sec:testing2}
We characterize uncertainty in our estimates using confidence intervals. A confidence interval (CI) of confidence level $(1-2\alpha)$ (with $0\leq\alpha\leq 1$) for an estimated parameter has the property that with probability $(1-2\alpha)$ when the experiment is performed and the CI calculated, the CI will contain the true value of the parameter. Bootstrap methods~\cite{Efron1} provide a way to calculate CIs, based on two steps: using simulations to build a set of estimates and applying an algorithm to the simulated estimates to produce the interval. For the first step, we use a ``parametric bootstrap'' in which estimates of $V_q$ and $V_p$ obtained from the original experiment are used to simulate experiments $N_B$ times according to the model described by Eq. (\ref{EQPnDodonov2}), producing $N_B$ pairs of simulated estimates of $V_q$ and $V_p$ (or equivalently $r$ and $\bar{n}$).

We first tested the Percentile Method \cite{Efron1,carpenter} to calculate our confidence. For an estimate $\hat{\theta}$ of some parameter $\theta$ and an ordered set of simulated bootstrap estimates. $B_R=   ({{\hat{\theta}}_1,{\hat{\theta}}_2,...,{\hat{\theta}}_{N_B}})$, the $(1-2\alpha)$ CI is $[\hat{\theta}_{l},\hat{\theta}_{m}]$, where
\begin{equation}
\begin{split}
&l =  \lfloor N_B \alpha\rfloor\\
&m = \lfloor N_B (1-\alpha) \rfloor.
\end{split}
\label{EQbinterval}
\end{equation}
We performed a test of the Percentile Method on three different reference states given by fixed $\bar{n} = 0.01$ and squeezing $r = (0, 1.0, 2.5)$. For each state, we ran 100 simulated experiments with $N = 10^4$. For each we calculated a $90\%$ confidence interval ($\alpha = 0.05$) and then estimated the coverage probabilities by the fraction of times that the confidence intervals contained the true value, with $V_q$, $V_p$, $r$, and $\bar{n}$ considered independently. For comparison we tested both $N_B=1,000$ and $2,000$.
Table \ref{tab:tabtestnbias} presents the results of the test. $N_B = 1,000$ gives good results for lower squeezing, but as squeezing increases the coverage probabilities decrease for all parameters but $V_p$. Doubling the number of bootstrap simulations does not significantly improve the coverage probabilities. For $r=2.5$, the coverage probability for $\bar{n}$ is far from expected. Coverage probabilities of around $98\%$ are obtained for estimates of $r$ when $r=0$ because this is on the boundary of parameter space and no bootstrap simulation can give estimates below $0$.

\begin{table*}
    \centering
    \begin{tabular}{ m{1.7cm} |  m{1.1cm} m{1.1cm} m{1.2cm} m{1.2cm} | m{1.4cm} m{1.4cm} m{1.4cm} m{1cm}}
\hline
State ($r$,$\bar{n}$) &   $V_p$(1k) &  $V_q$(1k) &  $r$(1k) &  $\bar{n}$(1k)&  $V_p$(2k) & $V_q$(2k) & $r$(2k) & $\bar{n}$(2k)\\
\hline
(0, 0.01) & 83\% & 88\% & 98\% & 94\% & 84\% & 89\% & 98\% & 93\%\\
(1.0, 0.01) & 82\% & 88\% & 87\% & 56\%& 83\% & 88\% & 86\% & 57\% \\
(2.5, 0.01) & 86\% & 58\% & 70\% & 15\% & 87\% & 59\% & 73\% & 18\% \\
\end{tabular}
    \caption{Table of the coverage probabilities for nominal 90\% confidence intervals using the percentile method. The coverage probabilities were estimated from 100 simulated experiments using $N_B=1000$ (left) and $N_B=2000$ (right) bootstrap replicates. The coverage probabilities significantly different from $90\%$ motivate our use of bias correction, shown in Table \ref{tab:tabtestwbias}.}
    \label{tab:tabtestnbias}
\end{table*}

To understand the low coverage probabilities we explore the ratio B$/\sigma$ of bias (here denoted by B) to standard deviation for estimates of parameters of different states, as shown in Table~\ref{tab:tabvpvq}. Bias is the difference between the expectation value of an estimate of a parameter and the true value of that parameter. We calculated B$/\sigma$ for both quadrature variances, $r$, and $\bar{n}$ for several states. We have previously seen that the highest squeezing considered, $r=2.5$, presented the worst results of average fidelity, and we can see in Table~\ref{tab:tabvpvq} that high squeezing also causes large B$/\sigma$ ratios. Such large B$/\sigma$ ratios, especially for estimates of $\bar{n}$ at high squeezing, could be the cause of the low coverage probabilities obtained when using the percentile method to construct confidence intervals.

\begin{table*}
    \centering
    \begin{tabular}{ m{1.1cm} | m{2cm} m{2cm} m{2cm} m{2cm} m{2cm} m{2cm}}
\hline
 True $r$ & B$/ \sigma$ in $r$ & True $V_p$ & B$/ \sigma$ in $V_p$ & True $V_q$ & B$ /\sigma$ in $V_q$ & B$/ \sigma$ in $\bar{n}$\\
\hline
0 & 0.69 & 0.51 & 0.68 & 0.51 & -0.68 & 0.069\\
0.5 & -0.16 & 1.39 & -0.19 & 0.19 & 0.013 & -0.25\\
1 & -0.24 & 3.77 & -0.32 & 0.069 & 0.16 & -0.64\\
1.5 & -0.020 & 10.24 & -0.12 & 0.025 & -0.085 & -0.90\\
2 & 0.27 & 27.85 & 0.094 & 0.0093 & -0.44 & -1.098\\
2.5 & 0.45 & 75.69 & 0.15 & 0.0034 & -0.72 & -1.40\\
\end{tabular}
    \caption{Estimates of bias $B$ divided by standard deviation $\sigma$ for estimates of parameters of states with various squeezing parameters $r$ and $\bar{n}=0.01$. Each estimate was generated from 1000 simulated experiments, containing $N=10000$ Fock measurements each, so some statistical fluctuation is expected. Estimates for which $|\text{B}/\sigma| > 1$ indicate that the bias in the estimate is significant when compared to the statistical uncertainty in the estimate, and so bias correction may be useful.}
    \label{tab:tabvpvq}
\end{table*}

To reduce the influence of bias we use the ``BC'' algorithm from \cite{efronbc} to calculate the CIs. The results are shown in Table \ref{tab:tabtestwbias} and were obtained using the same data used for Table \ref{tab:tabtestnbias}. The bias correction provides confidence intervals with coverage probabilities closer to the specified confidence level of $90\%$, though the coverage probability for $\bar{n}$ is still low.

\begin{table*}
    \centering
    \begin{tabular}{ m{1.7cm} |  m{1.1cm} m{1.1cm} m{1.2cm} m{1.2cm} | m{1.4cm} m{1.4cm} m{1.4cm} m{1cm}}
\hline
State (r,$\bar{n}$) &   $V_p$(1k) &  $V_q$(1k) &  r(1k) &  $\bar{n}$(1k)&  VP(2k) & VQ(2k) & r(2k) & $\bar{n}$(2k)\\
\hline
(0, 0.01) &  89\% & 88\% & 97\% & 97\% & 86\% & 90\% & 98\% & 97\%\\
(1.0, 0.01) & 87\% & 89\% & 88\% & 82\%& 89\% & 89\% & 89\% & 85\% \\
(2.5, 0.01) & 92\% & 89\% & 89\% & 74\% & 90\% & 86\% & 88\% & 77\% \\
\end{tabular}
    \caption{Table of the coverage probabilities for nominal 90\% confidence intervals with bias correction, computed for the same states and data as shown in Table \ref{tab:tabtestnbias}. We see the coverage probabilities significantly closer to 90\%, compared to the uncorrected intervals of Table \ref{tab:tabtestnbias}.}
    \label{tab:tabtestwbias}
\end{table*}

\begin{figure}
    \centering
    \includegraphics[scale=0.26]{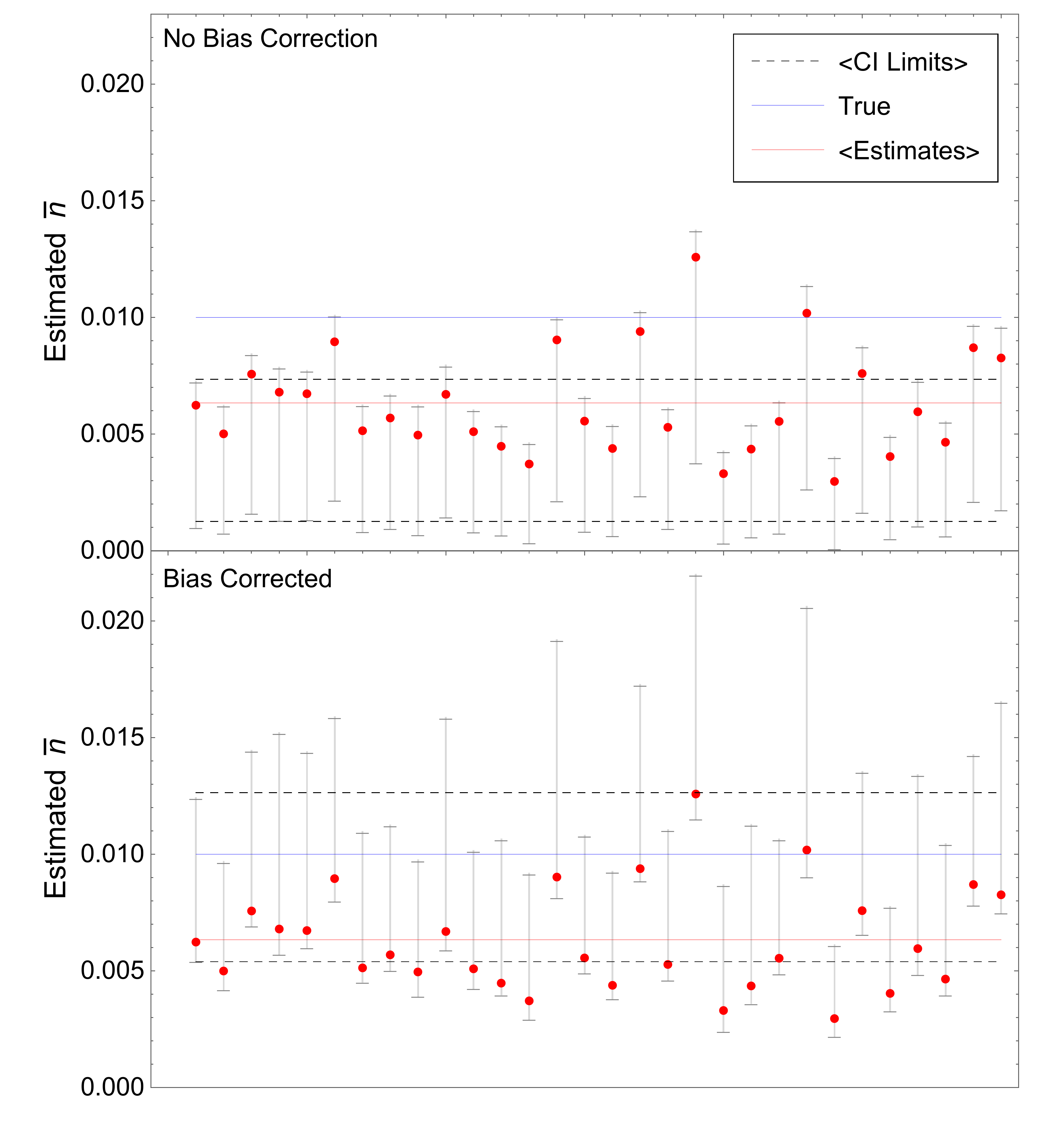}
    \caption{Example confidence intervals for $\bar{n}$, computed with the percentile method (upper panel) and the bias correcting BC method (lower panel) arranged in arbitrary order along the horizontal axis. The true parameters are $r = 2.5$ and $\bar{n} = 0.01$. The red dots show the point estimates for $\bar{n}$, the gray error bars show the confidence intervals with and without bias correction, the blue lines show the true value of $\bar{n}$, the red lines show the mean of the point estimates, and the dashed lines show the means of the upper and lower ends of the confidence intervals.
    \label{fig:intervals}}
\end{figure}

Figure \ref{fig:intervals} shows a set of 30 example 90\% intervals for $\bar{n}$ calculated for the tests presented on Tables~\ref{tab:tabtestnbias} and \ref{tab:tabtestwbias}.  After bias correction, far more of the confidence intervals contain the true value of $\bar{n}$.  For example, for the state with $r=2.5$ and $\bar{n}=0.01$, measured $N=1000$ times, when nominal $90 \%$ confidence intervals are calculated, the coverage probability for $\bar{n}$ has increased from $15 \%$ to $74 \%$ by using the bias correction. Because of the $\bar{n} \geq 0$ boundary, we expect that it will be difficult to achieve exact coverage probability for states with $\bar{n}$ near $0$ using standard methods, but given the bootstrap estimates, the bias correction is easy to apply and significantly improves the coverage probability.

\section{Conclusion}
\label{sec:conclusion}
In this work we presented a method of inference for any squeezed thermal state based on the weighted least-squares estimator. The results of our tests using simulated data for a squeezed thermal state showed high fidelity results. The use of bias-correcting confidence intervals mitigates the bias present in the point estimates from smaller data sets of highly-squeezed, low temperature states. These tools allow one to learn key properties of squeezed thermal states without the need for a phase reference (such as a local oscillator) in systems such as trapped ions~\cite{scburd} or integrated quantum optics~\cite{sahin2013, hopker2019}, where performing Fock measurements is convenient. An open problem that we hope to address in future work is the use of Fock populations to estimate the magnitude of displacement (from the origin of quadrature space) in addition to the squeezing and temperature of any single-mode Gaussian state.  We also would like to explore a hypothesis test for a measured Fock distribution being produced by a Gaussian state.

\section{Acknowledgments}
We thank Arik Avagyan, Shaun C. Burd, Hannah M. Knaack, and Emanuel Knill for helpful advice on the project, and we thank Arik Avagyan and Hannah M. Knaack for their editorial comments on the paper. I. P. B. thanks Universidade Estadual do Ceará (UECE) for financial support.  This work includes contributions of the National Institute of Standards and Technology, which are not subject to U.S. copyright.  The use of trade, product and software names is for informational purposes only and does not imply endorsement or recommendation by the U.S. government.

\bibliography{paper1.bbl}
\end{document}